\documentclass[]{spie}  

\usepackage{amsmath,amsfonts,amssymb}
\usepackage{graphicx}
\usepackage[colorlinks=true, allcolors=blue]{hyperref}

\newcommand\T{\rule{0pt}{2.6ex}}
\newcommand\B{\rule[-1.2ex]{0pt}{0pt}}

\title{Combining angular differential imaging and accurate polarimetry with SPHERE/IRDIS to characterize young giant exoplanets\footnote{\hspace{0.32cm}Observations were made with ESO Telescopes at the La Silla Paranal Observatory under program ID: 098.C-0790(A).}}

\author[1,a]{Rob G. van Holstein}
\author[a]{Frans Snik}
\author[b]{Julien H. Girard}
\author[a]{Jozua de Boer}
\author[a]{Christian Ginski}
\author[a]{Christoph U. Keller}
\author[c]{Daphne M. Stam}
\author[d]{Jean-Luc Beuzit}
\author[d]{David Mouillet}
\author[e]{Markus Kasper}
\author[f]{Maud Langlois}
\author[g,h]{Alice Zurlo}
\author[a]{Remco J. de Kok}
\author[i]{Arthur Vigan}

\affil[a]{Leiden Observatory, Leiden University, P.O. Box 9513, 2300 RA Leiden, The Netherlands} 
\affil[b]{Space Telescope Science Institute, 3700 San Martin Drive, Baltimore, MD 21218, USA}
\affil[c]{Faculty of Aerospace Engineering, Delft University of Technology, Kluyverweg 1, 2629 HS Delft, The Netherlands}
\affil[d]{CNRS, IPAG, F-38000 Grenoble, France}
\affil[e]{European Southern Observatory, Karl-Schwarzschild-Str. 2, 85748 Garching, Germany}
\affil[f]{Centre de Recherche Astrophysique de Lyon, CNRS/ENS-L/Université Lyon 1, 9 av. Ch. André, 69561 Saint-Genis-Laval, France}
\affil[g]{N\'ucleo de Astronom\'ia, Facultad de Ingenier\'ia, Universidad Diego Portales, Av. Ejercito 441, Santiago, Chile}
\affil[h]{Millennium Nucleus `Protoplanetary Disk', Departamento de Astronom\'ia, Universidad de Chile, Casilla 36-D, Santiago, Chile}
\affil[i]{Aix Marseille Univ, CNRS, LAM, Laboratoire d'Astrophysique de Marseille, Marseille, France}

\authorinfo{$^1$ E-mail: vanholstein@strw.leidenuniv.nl}

\begin{document} 
\maketitle

\begin{abstract}
Young giant exoplanets emit infrared radiation that can be linearly polarized up to several percent. This linear polarization can trace: 1) the presence of atmospheric cloud and haze layers, 2) spatial structure, e.g. cloud bands and rotational flattening, 3) the spin axis orientation and 4) particle sizes and cloud top pressure. We introduce a novel high-contrast imaging scheme that combines angular differential imaging (ADI) and accurate near-infrared polarimetry to characterize self-luminous giant exoplanets. We implemented this technique at VLT/SPHERE-IRDIS and developed the corresponding observing strategies, the polarization calibration and the data-reduction approaches. The combination of ADI and polarimetry is challenging, because the field rotation required for ADI negatively affects the polarimetric performance. By combining ADI and polarimetry we can characterize planets that can be directly imaged with a very high signal-to-noise ratio. We use the IRDIS pupil-tracking mode and combine ADI and principal component analysis to reduce speckle noise. We take advantage of IRDIS' dual-beam polarimetric mode to eliminate differential effects that severely limit the polarimetric sensitivity (flat-fielding errors, differential aberrations and seeing), and thus further suppress speckle noise. To correct for instrumental polarization effects, we apply a detailed Mueller matrix model that describes the telescope and instrument and that has an absolute polarimetric accuracy $\leq0.1\%$. Using this technique we have observed the planets of HR 8799 and the (sub-stellar) companion PZ Tel B. Unfortunately, we do not detect a polarization signal in a first analysis. We estimate preliminary $1\sigma$ upper limits on the degree of linear polarization of $\sim1\%$ and $\sim0.1\%$ for the planets of HR 8799 and PZ Tel B, respectively.
The achieved sub-percent sensitivity and accuracy show that our technique has great promise for characterizing exoplanets through direct-imaging polarimetry.
\end{abstract}


\section{INTRODUCTION}
\label{sec:intro}

Direct imaging enables the characterization of the atmospheres of self-luminous, hot, massive planets with photometry and spectroscopy. 
	With these methods, the planets' luminosity and atmospheric composition, structure and temperature can be constrained (see e.g.~Refs.~\citenum{lafreniere_exoplanet, ingraham_hr8799cd, chilcote_betapic, macintosh_erib, zurlo_hr8799}). 
%
%
Additional information on the composition and structure of planetary (or sub-stellar companion) atmospheres can be deduced with polarimetry. 

	Not only the starlight that an exoplanet reflects is expected to be linearly polarized~\cite{seager_cegp, stam_exo}, but also the thermal emission of a planet (or brown dwarf), as this radiation from inside the atmosphere will be scattered by cloud and haze particles on its way up~\cite{sengupta_browndwarf, dekok_exopol, stolker_exopol}. 
	If the companion is spherically symmetric, polarization signals from different parts on its surface will cancel each other and the integrated degree of linear polarization will be zero~\cite{dekok_exopol}.
	Therefore, for a net polarization signal to arise in the thermally emitted radiation, the companion must feature asymmetries such as equatorial flattening due to rapid rotation, patchy clouds or spots in the atmosphere~\cite{dekok_exopol}, or an obscuring moon~\cite{sengupta_exomoons} or a circumplanetary disk~\cite{stolker_exopol}. 
	The degree of linear polarization at near-infrared wavelengths of hot exoplanets featuring such asymmetries is expected to be generally larger than 0.1\% and could be up to several percent in some cases~\cite{dekok_exopol, stolker_exopol}.

Measurements of the polarized thermal emission of exoplanets could provide information on the presence and patchiness of atmospheric clouds and hazes, the cloud top pressure, spatial structure (the asymmetries mentioned above), and the surface gravity and mass of the companion~\cite{dekok_exopol, marley_exopol}. 
	To disentangle the various possible causes of the polarization, polarimetric follow-up observations are needed.
	By determining the angle of linear polarization, the planet's projected spin axis could be constrained~\cite{dekok_exopol}. 
	If the polarization signal is periodic, it could indicate the presence of persistent storms, such as Jupiter's Great Red Spot, and reveal atmospheric rotation rates.
	Finally, combining polarimetric measurements with flux measurements could reveal atmospheric particle properties, such as albedo and size. 
	The information on the atmospheric composition and structure revealed through polarimetry will significantly increase the accuracy of fitting atmospheric models based on known spectra of field brown dwarfs and sub-stellar companions to spectroscopic exoplanet observations, which currently results in errors of at least 10\%~\cite{ingraham_hr8799cd, chilcote_betapic, bonnefoy_hr8799}.	
																								

Near-infrared polarimetry has already been successfully performed for dozens of field brown dwarfs, yielding degrees of linear polarization between 0.1 to 2.5\% in the I-band~\cite{sengupta_browndwarf} and up to 0.8\% in the Z- and J-bands~\cite{milespaez_browndwarf}.
	For these field dwarfs, the polarization likely arises from patchy clouds.
 	The polarization signals of exoplanets are expected to be stronger, because exoplanets have a lower surface gravity, hence a stronger flattening for a given rotation rate, and a lower effective atmospheric temperature can yield stronger polarization signals for a given temperature gradient~\cite{dekok_exopol}. 
 	With the recently comissioned high-contrast imaging polarimeters SPHERE and GPI, detecting these polarization signals is now technically feasible. 
	Measurements of the thermal polarization signals of the planets of HR 8799 have already been attempted with VLT/NACO by Ref.~\citenum{juanovelar_thesis} and recently of HD 19467 B with GPI by Ref.~\citenum{jensen_padi}, but the contrasts attained were insufficient for a detection.
	The first direct measurement of exoplanetary polarization signals has therefore yet to be performed.	


SPHERE (Spectro-Polarimetric High-contrast Exoplanet REsearch) is a high-contrast imaging instrument employing extreme adaptive optics, stellar coronagraphs and three imagers to directly image and characterize giant exoplanets orbiting nearby stars~\cite{beuzit_sphere}. 
	It is installed on the Nasmyth platform of UT3 of the Very Large Telescope (VLT).
	One of the imagers, the InfraRed Dual-band Imager and Spectrograph (IRDIS)~\cite{dohlen_irdis} is primarily designed for detecting hot young exoplanets. 
	It also	has a dual-beam polarimetric mode that is mainly used for high-contrast imaging of circumstellar disks~\cite{langlois_irdis, deboer_irdis}.
	Since IRDIS has detected exoplanets with huge signal-to-noise ratio ($SNR > 200$ for some planets of HR 8799~\cite{zurlo_hr8799}), exoplanetary polarization signals could be detected for the first time with this polarimetric mode.
	Because the expected polarization signals are at most a small fraction ($<1\%$) of the total thermal signal of a planet, we need to achieve a \emph{polarimetric sensitivity}, i.e.~the noise level in the degree of linear polarization, and an \emph{absolute polarimetric accuracy}, i.e.~the uncertainty in the measured polarization signal, of $\sim0.1\%$. 

In an attempt at measuring polarization signals of sub-stellar companions, we have observed the four young giant planets HR 8799 bcde~\cite{marois_hr8799, marois_hr8799e} and the (sub-stellar) companion PZ Tel B~\cite{biller_pztel, mugrauer_pztel} with SPHERE/IRDIS.
	The planets of HR 8799 could be polarized, because recent spectral measurements have revealed sub-micron dust particles in their atmospheres~\cite{bonnefoy_hr8799}.
	In addition, temporal variations in near-infrared gaseous absorption features, such as those of CH$_4$, strongly suggest the presence of patchy clouds~\cite{oppenheimer_hr8799}.
	PZ Tel B is a very bright companion in an eccentric orbit seen edge-on~\cite{ginski_companionorbits, maire_pztel}. 
	If its spin axis is perpendicular to its orbital plane, polarization due to rotational flattening could be detected.

To achieve the required polarimetric sensitivity and accuracy, 	we combine angular differential imaging (ADI)~\cite{marois_adi} with IRDIS' dual-beam polarimetric mode (see~Ref.~\citenum{juanovelar_thesis}) 
and correct for instrumental polarization effects with the detailed Mueller matrix model described in Ref.~\citenum{vanholstein_irdismodel}.
	In comparison to the recent attempt to measure near-infrared exoplanetary polarization with GPI~\cite{jensen_padi}, we observe longer time sequences to take full advantage of ADI and apply more advanced ADI and polarimetric demodulation techniques.
	In Secs.~\ref{sec:measurement_technique} and \ref{sec:observations}, the measurement technique and observations will be described, respectively. 
	Subsequently, the data-reduction scheme we developed for the observations will be discussed in Sec.~\ref{sec:data_reduction}.
	Section~\ref{sec:results} will then present the results and Sec.~\ref{sec:discussion} will discuss the measurement technique and possible improvements to the data-reduction.
	Finally, conclusions will be presented in Sec.~\ref{sec:conclusions}.

\section{MEASUREMENT TECHNIQUE}
\label{sec:measurement_technique}

The polarization signals of young giant exoplanets are expected to be a few tenths of a percent to a percent of the total intensity of the planets.
	To detect these polarization signals, we need to observe planets that can be directly imaged with a very high signal-to-noise ratio.
	IRDIS uses a (non-polarizing) beamsplitter and a pair of polarizers with orthogonal transmission axes to simultaneously create two adjacent images on the detector~\cite{langlois_irdis, deboer_irdis}.
	This dual-beam system allows us to perform beam switching with the half wave plate (HWP) and to compute the Stokes parameters $Q$ and $U$ from the double difference~\cite{bagnulo_spectropolarimetry} (see Sec.~\ref{sec:data_reduction_polarization}), thereby eliminating differential effects that severely limit polarimetric sensitivity, e.g.~flat-fielding errors, differential aberrations and seeing (see e.g.~Ref.~\citenum{canovas_data}).
	To attain the high contrast required for polarimetry of exoplanets, we observe in pupil-tracking mode and construct both the total intensity and polarization images by combining ADI with principal component analysis~\cite{amara_pynpoint, soummer_klip, meshkat_opca} (PCA) to significantly reduce speckle noise, the principle noise component. 
	The combination of ADI and polarimetry further suppresses speckle noise (especially at small angular separations), because stars, and therefore the speckles, are generally unpolarized. 

To accurately derive a planet's polarization state from a measurement,
it is paramount to correct for instrumental polarization (IP) and cross-talk of the complete optical system, i.e.~telescope and instrument.
	To this end, we use the Mueller matrix model for VLT/SPHERE/IRDIS described in Ref.~\citenum{vanholstein_irdismodel}.
	This model has been validated by taking measurements of an unpolarized standard star and with SPHERE's internal source, reaching an absolute polarimetric accuracy $\leq0.1\%$ in all broadband filters (Y-, J-, H- and K$_\mathrm{s}$-band; see Tab.~\ref{tab:bb_filters}).
	The polarimetric accuracy is particularly affected by the IP, which can make unpolarized sources appear a few percent polarized when unaccounted for.
	With the model, the IP can be subtracted more accurately than with regular IP-subtraction techniques, because no assumptions on the stellar polarization are needed.

To enable the application of ADI + PCA, we have commissioned pupil-tracking for IRDIS' polarimetric mode. 
	In this mode, the derotator (K-mirror) only compensates for the altitude angle of the telescope, so that the pupil (and the quasi-static speckle pattern) is kept fixed with respect to the detector, while the image (the planet) rotates with the parallactic angle.
	Unfortunately we could not implement a new HWP rotation control law to keep the polarization direction of the source fixed on the detector during pupil-tracking\footnote{For this, the following HWP control law should be implemented:
\begin{equation}
	\theta_{\mathrm{HWP}} = a + \frac{1}{2}\left(-p + \gamma + \eta_\mathrm{pupil}\right)
	\label{eq:hwp_law_pt}
\end{equation} 
with $\theta_{\mathrm{HWP}}$ the HWP angle, $a$ the altitude angle of the telescope, $p$ the parallactic angle of the target, $\gamma$ an offset of the angle of linear polarization due to a user-defined HWP offset and $\eta_\mathrm{pupil}$ the fixed position angle offset of the image to align the `spider mask' with the diffraction pattern of the support structure of the telescope's secondary mirror (see below). Beam switching with the HWP (to measure Stokes $Q$ and $U$) is performed relative to this HWP angle. Reference~\citenum{vanholstein_irdismodel} describes the HWP control law that is used in field-tracking mode.}.
	Therefore Stokes $Q$ and $U$ are measured by performing beam switching with the HWP relative to the vertical (perpendicular to the Nasmyth platform; STATIC mode) with HWP switch angles $0^\circ$ and $45^\circ$ to measure $Q$, and $22.5^\circ$ and $67.5^\circ$ to measure $U$ (a so-called polarimetric or HWP cycle).
	The disadvantage is that the polarization direction of the source rotates with the parallactic angle on the detector while tracking a target, but this can readily be accounted for with the Mueller matrix model of Ref.~\citenum{vanholstein_irdismodel}.
	Pupil-tracking with IRDIS' polarimetric mode is officially offered since P100.
\begin{table}
\centering
\caption{Central wavelength and bandwidth of the broadband filters available for IRDIS polarimetry~\protect\cite{sphere_manual}.} 
\centering 
\begin{tabular}{l c c} 
\hline\hline
\T\B Filter & Central wavelength (nm) & Bandwidth (nm) \\ 
\hline
\T BB\_Y & 1042.5 & 139 \\
BB\_J & 1257.5 & 197 \\
BB\_H & 1625.5 & 291 \\
\B BB\_$\mathrm{K_s}$ & 2181.3 & 313.5 \\
\hline 
\end{tabular}  
\label{tab:bb_filters}
\end{table} 

\begin{figure}[!hbtp] 
\centering 
\includegraphics[width=12cm]{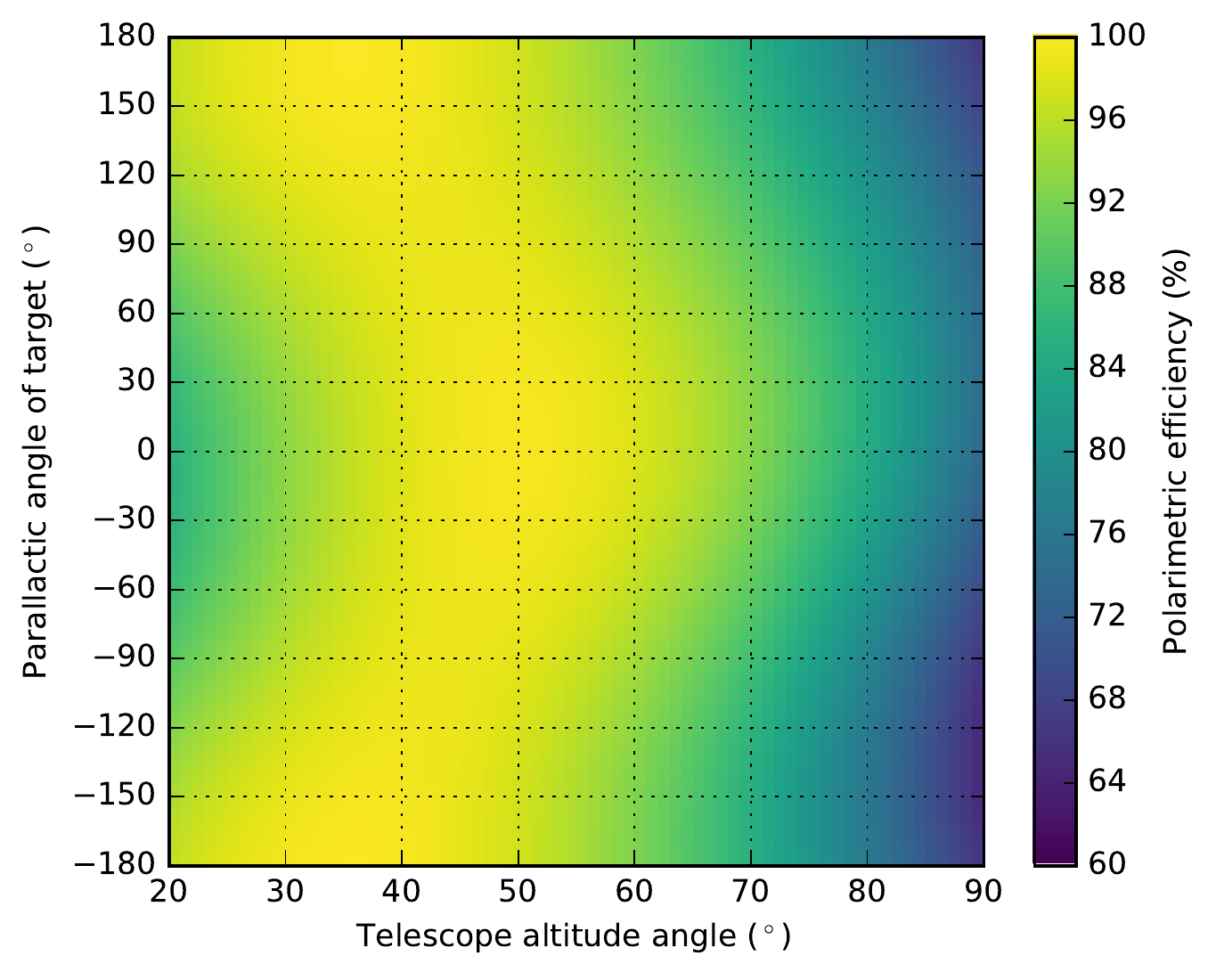} 
\caption{Polarimetric efficiency of SPHERE/IRDIS in H-band as a function of parallactic angle of the target and telescope altitude angle in pupil-tracking mode.} 
\label{fig:efficiency_pupil_tracking} 
\end{figure}

As described in Refs.~\citenum{vanholstein_irdismodel} and \citenum{deboer_irdis}, the derotator can produce very strong cross-talk at specific derotator angles, resulting in severe loss of polarization signal.
	For field-tracking observations, Ref.~\citenum{deboer_irdis} recommends to apply an offset to the derotator angle to prevent this signal loss.
	Such an offset cannot be applied in pupil-tracking mode, because the support structure of the secondary mirror of the telescope (the `spider') will then not be aligned with a mask added to the Lyot stop (the `spider mask'), resulting in (locally) much higher speckle noise.	
Fortunately, the polarimetric efficiency, i.e.~the fraction of the linearly polarized light entering the system that is actually measured, happens to be sufficiently high for a large range of altitude angles in all broadband filters.
	Figure~\ref{fig:efficiency_pupil_tracking} shows the polarimetric efficiency in pupil-tracking mode as a function of parallactic and altitude angle in H-band and is constructed using the Mueller matrix model of Ref.~\citenum{vanholstein_irdismodel}.
	The efficiency is $>80\%$ for altitude angles between $20^\circ$ and $75^\circ$, but goes down to 64\% for altitude angles larger than $75^\circ$.
	The polarimetric efficiency plots in Y-, J-, and K$_\mathrm{s}$-band look similar, but with the minima at altitude angles between $20^\circ$ and $75^\circ$ equal to 87\%, 96\% and 83\%, respectively. At altitude angles larger than $75^\circ$, the minima are equal to 76\%, 94\% and 67\%, respectively.

\section{OBSERVATIONS}
\label{sec:observations}

During the nights starting on October 10 and 12, 2016, we observed the four young giant planets HR 8799 bcde and the (sub-stellar) companion PZ Tel B with the measurement technique described in Sec.~\ref{sec:measurement_technique}. 
	An overview of the observations is shown in Tab.~\ref{tab:observations}. 
	PZ Tel was observed twice for almost 40 min during twilight using the broadband H- and J-band filters, while HR 8799 was observed twice for $\sim2.5$ h during nighttime in H-band.
	All measurements were taken with the apodized pupil Lyot coronagraph ALC\_YJH\_S (mask diameter = 185 mas).
	The first three observing sequences had good to medium seeing conditions.
	However, the HR 8799 observations of 13-10-2016 are not considered in this work, as the seeing and coherence time were very poor and the control loop of the adaptive optics system opened many times.
	PZ Tel and HR 8799 were observed at an average altitude angle of $\sim57^\circ$ (air mass $\sim1.2$) and $\sim41^\circ$ (air mass $\sim1.5$), respectively.
	From Fig.~\ref{fig:efficiency_pupil_tracking}, it follows that the polarimetric efficiency of the observations in H-band at these altitude angles is always higher than $\sim86\%$.
\begin{table}[!hbtp] 
\caption{Overview of the polarimetric observations of the (sub-stellar) companion PZ Tel B and the four giant planets around HR 8799. DIT = Detector Integration Time; NDIT = Number of Detector Integration Times; ToT = Time on Target; $m_\mathrm{V}$ = apparent magnitude of central star in V-band.} 
\centering 
\resizebox{\textwidth}{!}{%
\begin{tabular}{cllcccrcccll} 
\hline
Date & Target & Filter & DIT (s) & \begin{tabular}{@{}c@{}} HWP \T \\ cycles \B \end{tabular} & NDIT & \begin{tabular}{@{}c@{}} ToT \T \\ (min) \B \end{tabular} & \begin{tabular}{@{}c@{}} Parallactic \T \\ rotation ($^\circ$) \B \end{tabular} & \begin{tabular}{@{}c@{}} Altitude \T \\ angle ($^\circ$) \B \end{tabular} & $m_\mathrm{V}$ & Seeing ($^{\prime\prime}$) & \begin{tabular}{@{}c@{}} Coherence \T \\ time (ms) \B \end{tabular} \T\B \\ 
\hline 
10-10-2016 & PZ Tel & BB\_H & 12 & 10 & 4 & 39 & 14 & 55 - 60 & 8.34 & 0.6 - 1.0 & 2.8 - 6.4 \T \\
11-10-2016 & HR 8799 & BB\_H & 16 & 43 & 3 & 164 & 50 & 38 - 44 & 5.95 & 0.4 - 0.9 & 2.4 - 6.0 \\
12-10-2016 & PZ Tel & BB\_J & 12 & 10 & 4 & 37 & 13 & 53 - 58 & 8.34 & 0.9 - 1.2 & 2.8 - 6.1 \\
13-10-2016 & HR 8799 & BB\_H & 16 & 39 & 3 & 145 & 45 & 40 - 44 & 5.95 & 0.9 - 2.8 & 1.0 - 4.0 \B \\ 
\hline 
\end{tabular}}
\label{tab:observations} 
\end{table} 

\section{DATA-REDUCTION}
\label{sec:data_reduction}

	To explain the data-reduction, we will primarily discuss the reduction of the HR 8799 data of 11-10-2016 and only mention the details of the reduction of the PZ Tel datasets that deviate from this first description.
	We prepare the raw frames by applying dark-subtraction, flat-fielding and bad pixel filtering.
	Subsequently, we center the frames with the four satellite spots of the star center frames (see~Ref.~\citenum{zurlo_hr8799}).
	From these prepared frames, we create images of Stokes $Q$ and $U$ and the corresponding total intensity images $I_Q$ and $I_U$ to determine the degree and angle of linear polarization of the planets.
	We will now first describe the creation of the total intensity images.

\subsection{Construction of Total Intensity \boldmath{$I_Q$}- and \boldmath{$I_U$}-images}

\begin{figure}[!hbtp] 
\centering 
\includegraphics[width=12cm]{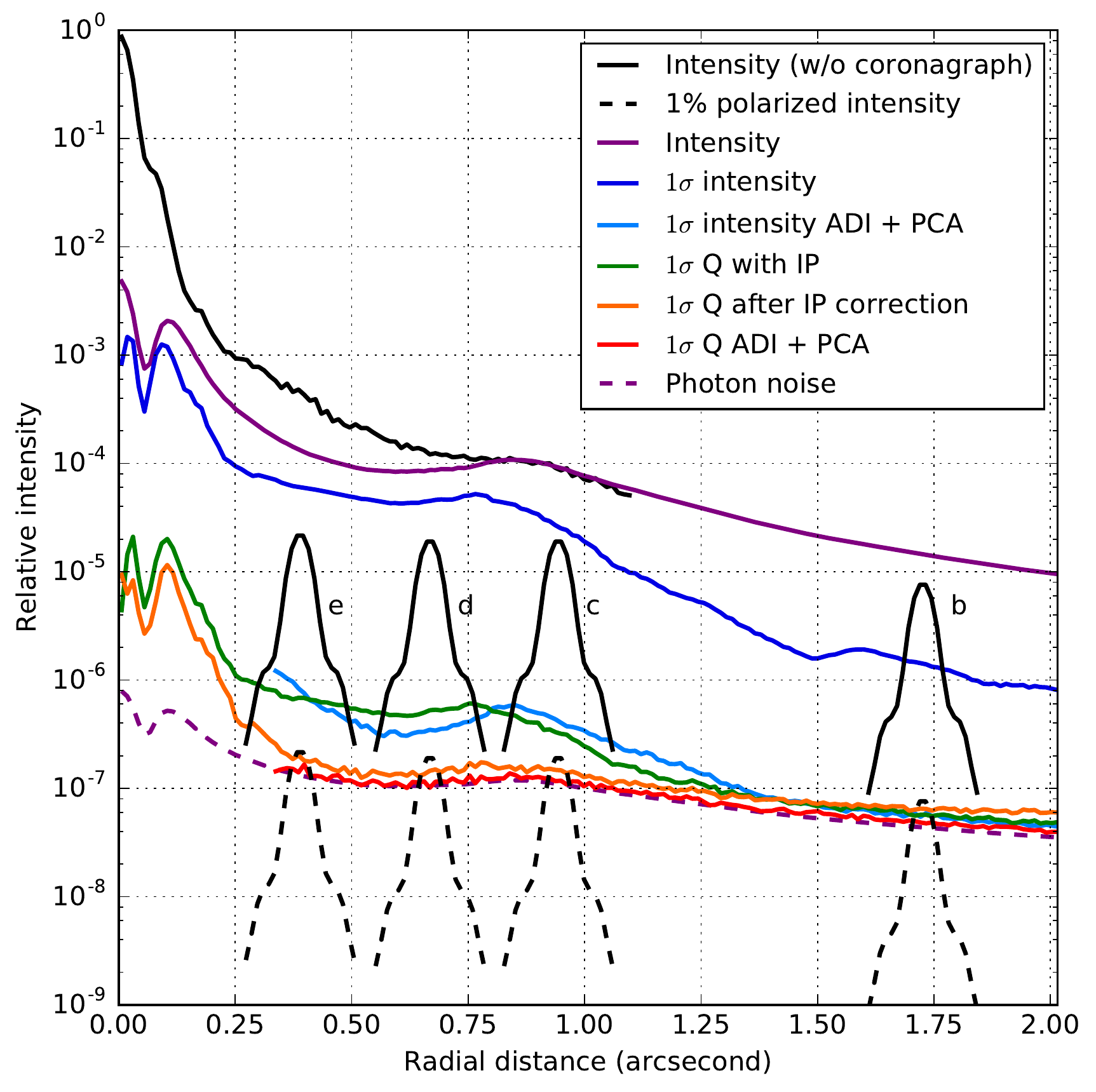} 
\caption{Contrast curve of the HR 8799 observations of 11-10-2016 in H-band showing the effect of the various data-reduction steps.} 
\label{fig:contrast_curve_hr8799} 
\end{figure}

To create the total intensity images, we start by adding the images of the simultaneously recorded orthogonal polarization states on the left and right halves of the detector (single sum images $I_Q^+$, $I_Q^-$, $I_U^+$ and $I_U^-$) for all the prepared frames. 
	After that, for each HWP cycle, we compute the double sum, i.e. the mean of the single sum images with HWP switch angles $0^\circ$ 
	and $45^\circ$ 
	for $I_Q$ and $22.5^\circ$ 
	and $67.5^\circ$ 
	for $I_U$.
	To show the sensitivity we attain, we look at the contrast curve displayed in Fig~\ref{fig:contrast_curve_hr8799}.
	The curves in this Figure are computed as the mean or $1\times$ the standard deviation ($1\sigma$) over 1 pixel wide annuli centered on the star, normalized with the maximum intensity of the star.	
	The purple solid curve in Fig.~\ref{fig:contrast_curve_hr8799} shows the mean intensity profile of the coronagraphic stellar point spread function (PSF) when we compute the mean of the cube of the $I_Q$-frames. 
	Comparing this curve to the non-coronagraphic stellar PSF profile (top black solid curve) obtained from the flux calibration image (see~Ref.~\citenum{zurlo_hr8799}) clearly demonstrates the effect of the adaptive optics and coronagraph close to the star. 
	The dark blue solid curve shows the standard deviation ($1\sigma$) over the $I_Q$-frames, which can be regarded as a measure of the speckle noise.
	It follows that we can detect planet b from the double sum images, but that the inner three planets remain hidden.
	Indeed, planet b is (marginally) visible in a single raw frame.
%

\begin{figure}[!hbtp] 
\centering 
\includegraphics[width=\hsize]{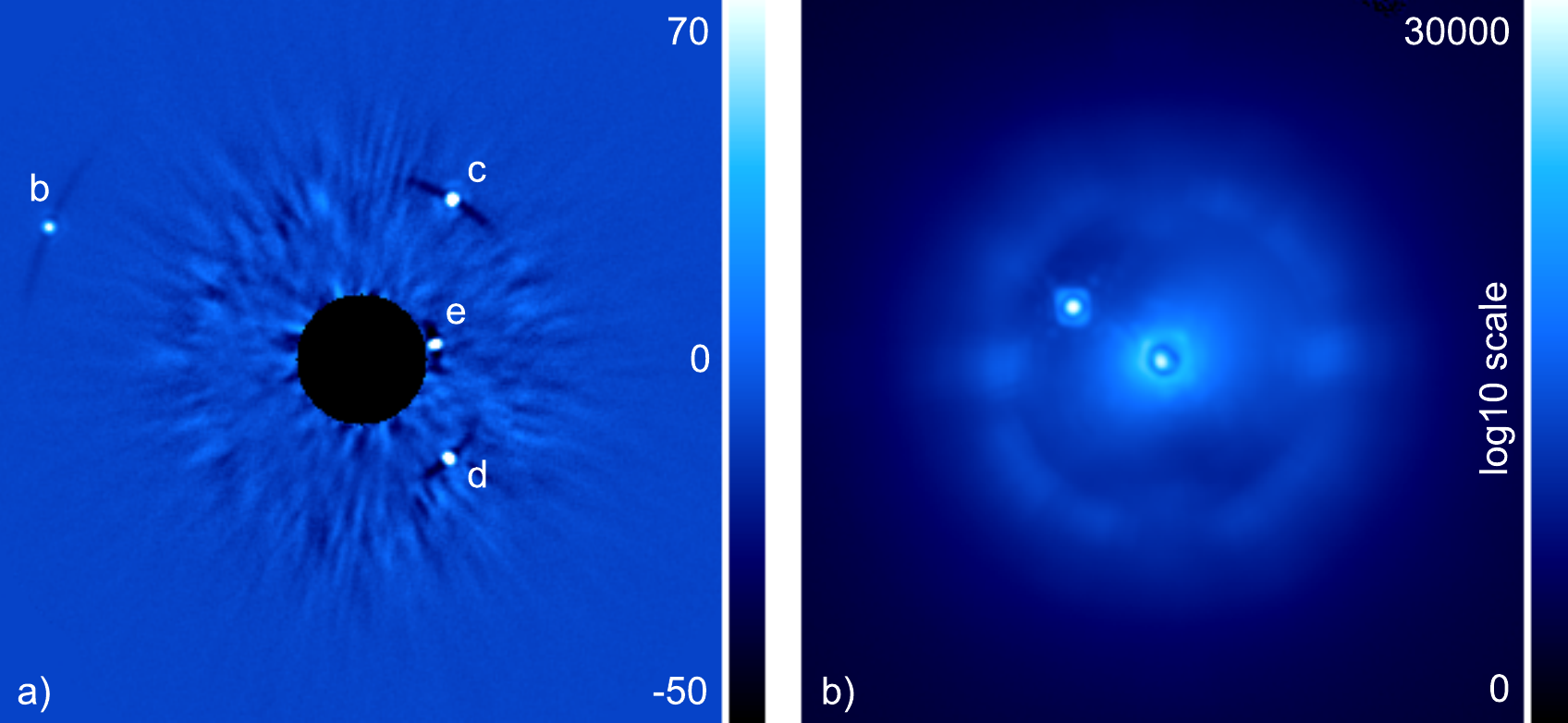} 
\caption{a) Total intensity $I_Q$-image of the planets of HR 8799 in H-band created using ADI + PCA. b) Total intensity $I_Q$-image of PZ Tel B in H-band (without ADI + PCA).} 
\label{fig:total_intensity_hr8799_pztel} 
\end{figure}

To increase the contrast, we combine ADI with principle component analysis (PCA), using the software package PynPoint~\cite{amara_pynpoint, amara_pynpointcode}, to subtract 5 principle components from the cubes of the $I_Q$- and $I_U$-frames separately.
	We then derotate the frames with the parallactic angle and compute their mean to obtain the final $I_Q$- and $I_U$-images.
	The resulting $I_Q$-image is shown in Fig.~\ref{fig:total_intensity_hr8799_pztel}a.
	The light blue solid curve in Fig.~\ref{fig:contrast_curve_hr8799} shows the standard deviation over this image and the four lower black solid curves show the PSF profiles of the planets\footnote{The PSF profiles of the planets are not extracted from the final $I_Q$-image, because we have not accounted for the ADI self-subtraction, for example by injecting fake negative planets~\cite{marois_photometry, bonnefoy_betapictorisb, zurlo_irdisphotometry}.
	Instead, we obtained them by scaling the non-coronagraphic stellar PSF of the flux calibration image with the planet contrasts reported in Ref~\citenum{zurlo_hr8799}.}.
	After applying ADI + PCA, the speckle noise is substantially suppressed and all planets are clearly detected. 
	This is also evident from Figure~\ref{fig:total_intensity_hr8799_pztel}a, as all planets and the Airy rings surrounding planet b and c (and perhaps d) are clearly visible.	
%

\begin{figure}[!hbtp] 
\centering 
\includegraphics[width=12cm]{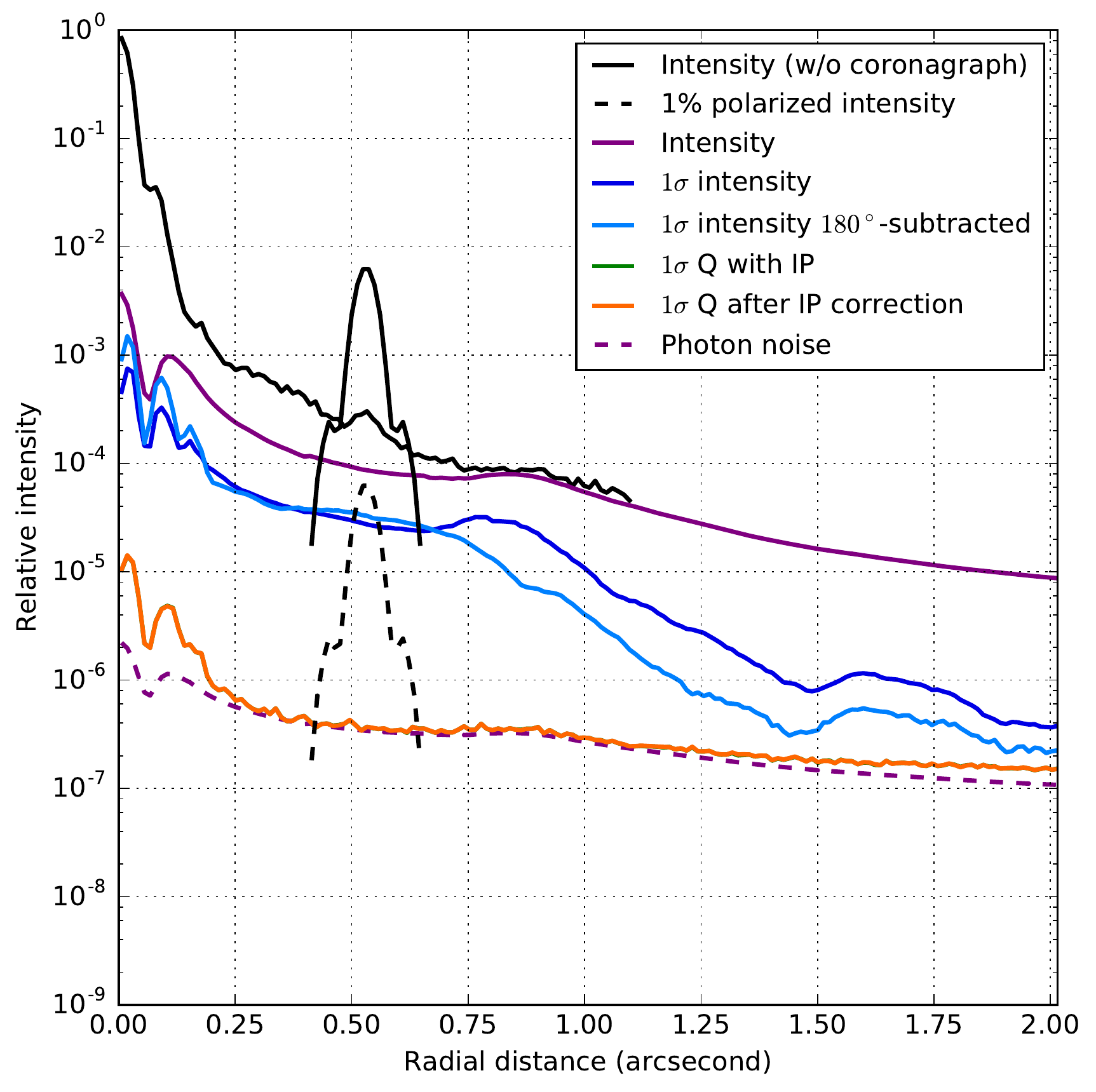} 
\caption{Contrast curve of the PZ Tel observations of 10-10-2016 in H-band showing the effect of the various data-reduction steps. The orange curve overlaps with the green curve, because the average of the IP in Stokes $Q$ happened to be zero.} 
\label{fig:contrast_curve_pztel} 
\end{figure}

For the reduction of the PZ Tel data, we do not perform ADI + PCA when creating the total intensity images, because the parallactic rotation is limited (see Tab.~\ref{tab:observations}).
	Instead we derotate the double sum $I_Q$- and $I_U$-frames with the parallactic angle and compute the mean of the cube of these derotated frames.
	The resulting $I_Q$-image of the H-band observations is displayed in Figure~\ref{fig:total_intensity_hr8799_pztel}b.
	The first Airy ring and the diffraction pattern of the support structure of the telescope's secondary mirror are clearly visible around the companion.
	To suppress the halo of starlight, which shows radial symmetry, we subtract $180^\circ$-rotated versions of the $I_Q$- and $I_U$-images from the originals.	
	Remaining background will be removed when computing the total intensity of the companion from apertures.
	The contrast curve of the PZ Tel observations in H-band is shown in Fig.~\ref{fig:contrast_curve_pztel}. 
	The PSF profile of the companion (lower solid black curve) is obtained from the halo-subtracted images.	
	The light blue solid curve in Fig.~\ref{fig:contrast_curve_pztel} shows that the companion is detected very clearly.

\subsection{Construction of Stokes \boldmath{$Q$}- and \boldmath{$U$}-images}
\label{sec:data_reduction_polarization}

\begin{figure}[!hbtp] 
\centering 
\includegraphics[width=\hsize]{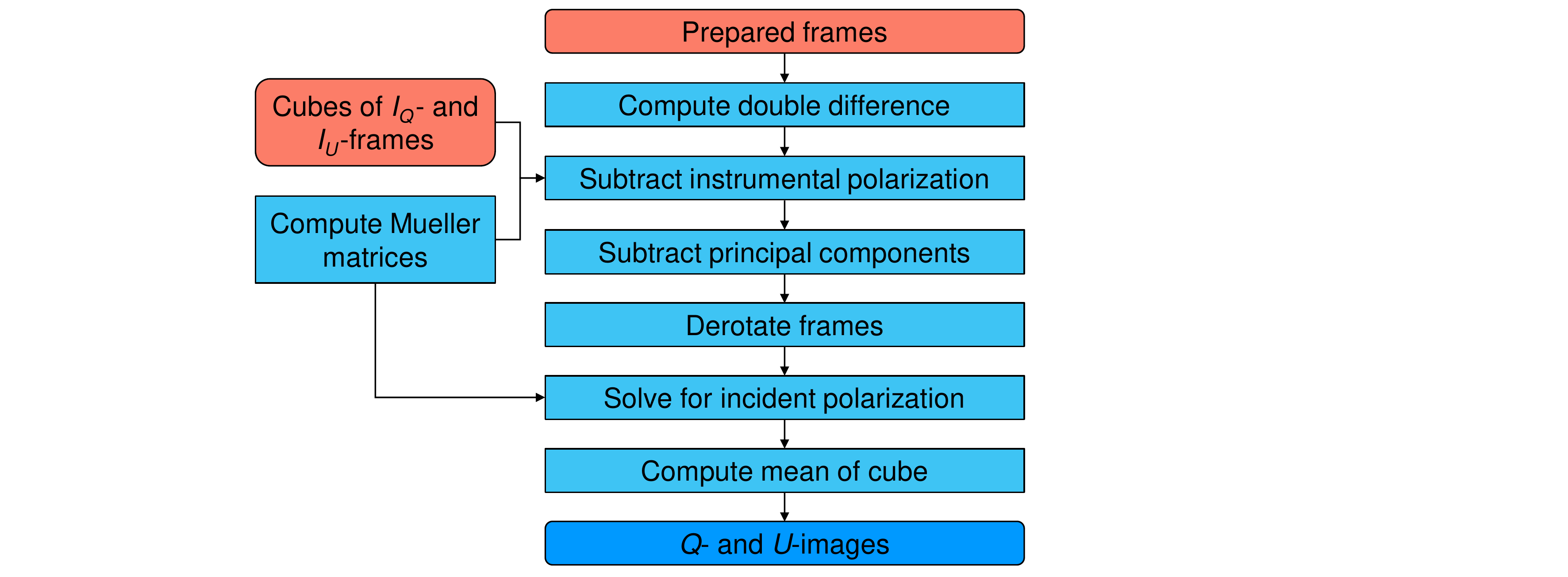} 
\caption{Flow diagram of the data-reduction steps to construct the images of Stokes $Q$ and $U$.} 
\label{fig:flow_chart} 
\end{figure}

In the contrast curve of the HR 8799 observations shown in Fig.~\ref{fig:contrast_curve_hr8799}, the tops of the planets' PSF profiles (four lower black solid curves) are at a contrast of $\sim10^{-5}$.
	If these planets are 1\% polarized, their polarization signals will correspond to the four black dashed curves.
	This clearly shows the challenge at hand: since the planets are expected to be at most a few tenths of a percent to a percent polarized, we need to reach a contrast in polarized light of at least $10^{-7}$. 
	In the following we will present and justify the order of data-reduction steps 
	to construct the Stokes $Q$- and $U$-images.
	A flow diagram of the data-reduction steps is shown in Fig.~\ref{fig:flow_chart}.

We start by computing the difference between the images of simultaneously recorded orthogonal polarization states on the left and right halves of the detector (single difference images $Q^+$, $Q^-$, $U^+$ and $U^-$) for all the prepared frames.
	For each HWP cycle, we then compute the double difference, i.e. half the difference between the single difference images with HWP switch angles $0^\circ$ and $45^\circ$ for $Q$, and $22.5^\circ$ and $67.5^\circ$ for $U$.
	We choose not to derotate the frames with the parallactic angle before computing the double difference, because this way we most effectively remove the quasi-static speckles and the IP downstream of the HWP, and suppress differential effects that limit the polarimetric sensitivity, including flat-fielding errors.
	The disadvantage is that we create some spurious polarization signal, because the planet positions have changed slightly between subsequent exposures.
	However, this rotation-induced signal cancels out when we integrate the polarization signal over a sufficiently large aperture.
	The double difference frames show a weak detector read-out artifact that has a continuous vertical band structure. 
	Similarly to Ref.~\citenum{deboer_diskcandidates}, we remove this structure by subtracting, for every column of pixels, the median value of the top and bottom 60 pixels.
	Since starlight shows little to no polarization (in general), computing the double difference strongly suppresses the halo of starlight.
	The green solid curve in Fig.~\ref{fig:contrast_curve_hr8799} shows that we reach approximately the same contrast as that attained using ADI + PCA on the total intensity images.

\begin{figure}[!hbtp] 
\centering 
\includegraphics[width=15cm]{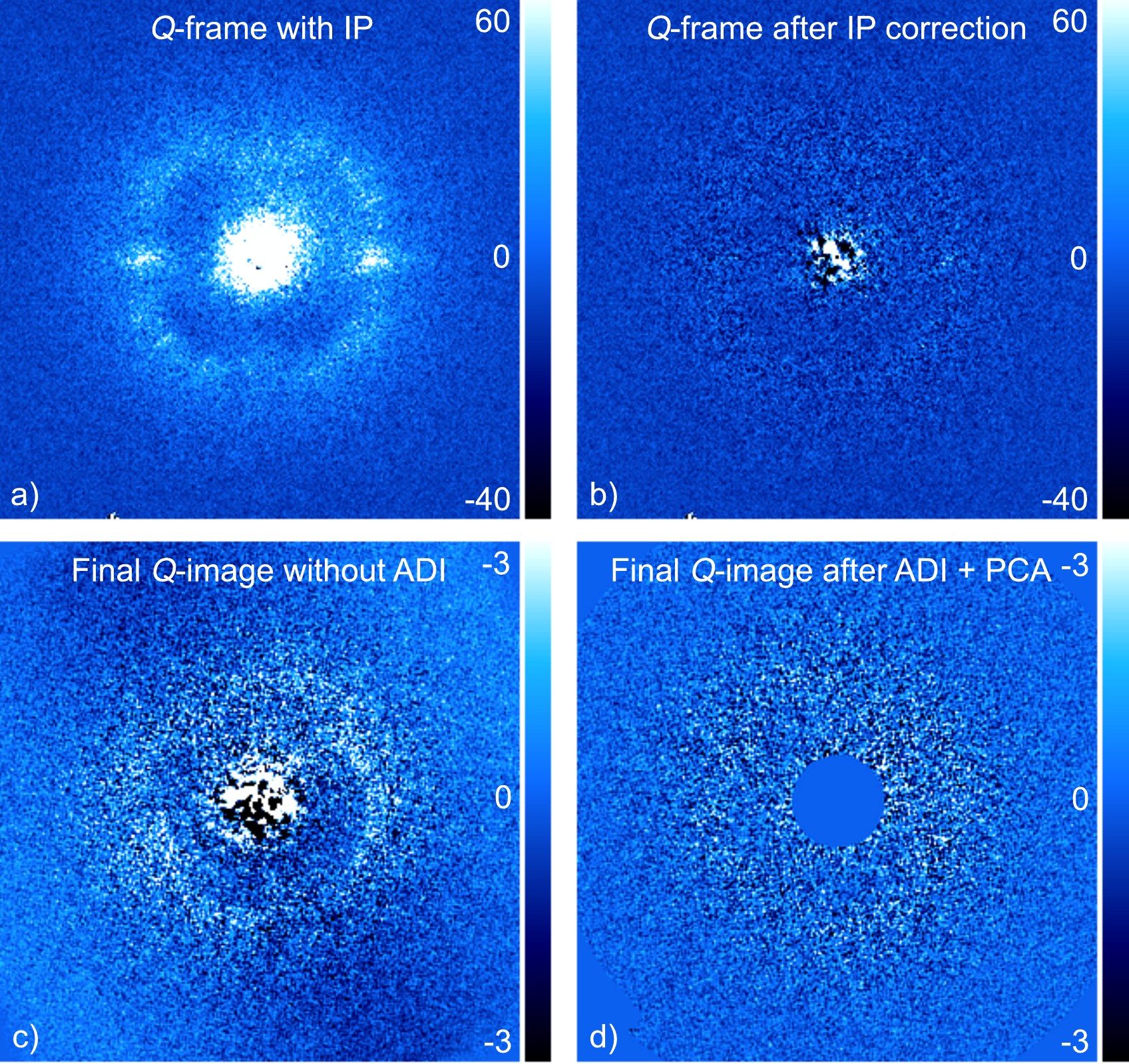} 
\caption{a) Double difference $Q$-frame showing instrumental polarization (IP) of SPHERE's first mirror (M4) and the telescope. b) $Q$-frame after removing the IP by subtracting the corresponding $I_Q$-frame scaled with the IP predicted by the Mueller matrix model. c) $Q$-image after computing the mean of the cube of IP-corrected $Q$-frames showing remaining structure. Note that the range of values is much smaller than that in frames a) and b). d) Final $Q$-image after removing the structure by subtracting three principle components and correcting for efficiency and cross-talk with the Mueller matrix model.} 
\label{fig:data_reduction_steps_Q_U} 
\end{figure}

\begin{figure}[!hbtp] 
\centering 
\includegraphics[width=\hsize]{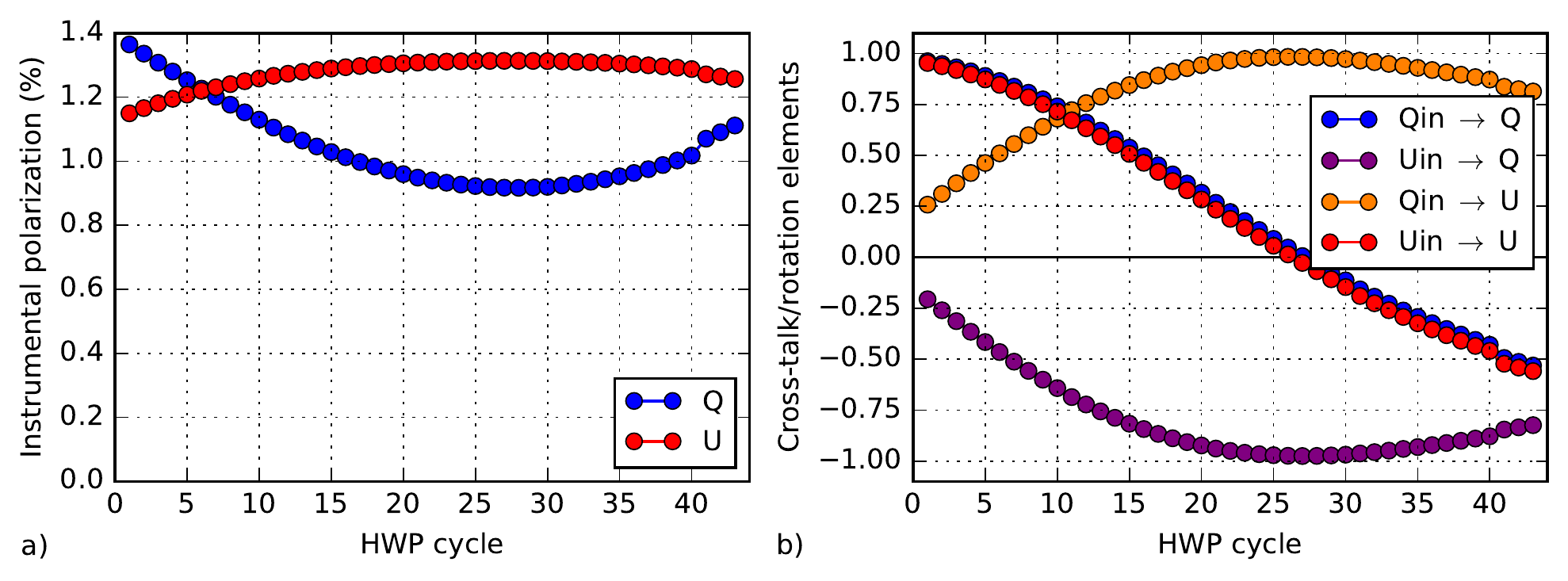} 
\caption{a) Instrumental polarization predicted by the Mueller matrix model for each double difference $Q$- or $U$-frame. b) The four elements describing the transmission ($Q_\mathrm{in} \rightarrow Q$ and $U_\mathrm{in} \rightarrow U$) and exchange between the polarization components ($U_\mathrm{in} \rightarrow Q$ and $Q_\mathrm{in} \rightarrow U$) as predicted by the Mueller matrix model for each pair of $Q$- and $U$-frames. Because we briefly interrupted the observing sequence to take calibration measurements, there is a small discontinuity after the 40th HWP cycle in both plots.} 
\label{fig:ip_retardance_hwp_cycle} 
\end{figure}

The IP created by SPHERE's first mirror (M4) and the telescope are not removed by computing the double difference, because they are located upstream of the HWP.
	Therefore a residual stellar speckle halo is visible in all double difference frames as shown in Fig.~\ref{fig:data_reduction_steps_Q_U}a.
	To remove this residual speckle halo and correct the planets' polarization signals for the IP, we use the Mueller matrix model of Ref.~\citenum{vanholstein_irdismodel} to describe the contribution of the telescope and instrument for every measurement.
	With the model we compute the IP for each double difference $Q$- or $U$-frame from the corresponding parallactic, telescope altitude, HWP and derotator angle.	
	The IP predicted by the model is shown in Fig.~\ref{fig:ip_retardance_hwp_cycle}a.
	We remove the IP in the $Q$- and $U$-frames by scaling their corresponding double sum intensity $I_Q$- and $I_U$-frames with the predicted IP and subtracting the result from the $Q$- and $U$-frames.
	The removed residual speckle halo in Fig.~\ref{fig:data_reduction_steps_Q_U}b illustrates that this procedure effectively removes close to all IP from the $Q$- and $U$-frames.
	As a result, the orange solid curve in Fig.~\ref{fig:contrast_curve_hr8799} shows a significant increase in contrast close to the star.

When we compute the mean of the cubes of the IP-corrected $Q$- and $U$-frames, some structure remains, as shown in Fig.~\ref{fig:data_reduction_steps_Q_U}c.
	This structure likely arises because the speckle field changes in time and the orthogonal polarization states are treated slightly differently by the instrument.
	In addition, there might be some stellar and background polarization or residual IP, because the Mueller matrix model has a finite accuracy.
	To remove the structure, we apply ADI + PCA to the cubes of the IP-corrected $Q$- and $U$-frames separately, subtracting three principle components for each.
	The ADI + PCA step cannot be performed before removing the IP with the Mueller matrix model, because in that case the IP will be partly removed with ADI + PCA and then over-subtracted with the modeled IP value.
	Also, applying ADI + PCA alone would never suffice to remove the IP at the planet position, because it cannot discriminate between planet signal and IP and the planets are moving between exposures.
	
After performing ADI + PCA, we derotate the frames so that the images are aligned with North up and the planets are at the same position in all frames.
	Since the HWP control law for pupil-tracking has not yet been implemented (see Sec.~\ref{sec:measurement_technique}), the polarization direction of the planets on the detector rotates between HWP cycles.
	In addition, the cross-talk produced by the instrument (in particular by the derotator) results in a time-varying transmission of and exchange between the polarization components $Q$ and $U$.
	As a result, the polarization signal of the planets is different in each pair of $Q$- and $U$-frames and we cannot simply compute the mean of the cubes of the $Q$- and $U$-frames.
	To correct for the rotation and the cross-talk, we use the Mueller matrix model to derive for each pair of $Q$- and $U$-frames (each HWP cycle) two linear equations describing the measurements.
	For every pixel, we solve these equations for the polarization signal incident on the telescope ($Q_\mathrm{in}$ and $U_\mathrm{in}$).
	The four elements describing the transmission ($Q_\mathrm{in} \rightarrow Q$ and $U_\mathrm{in} \rightarrow U$) and exchange between the polarization components ($U_\mathrm{in} \rightarrow Q$ and $Q_\mathrm{in} \rightarrow U$) are plotted in Fig.~\ref{fig:ip_retardance_hwp_cycle}b.
	Finally, we compute the mean of the cubes of the $Q_\mathrm{in}$- and $U_\mathrm{in}$-frames to obtain the final $Q$ and $U$-images incident on the telescope.

Figure~\ref{fig:data_reduction_steps_Q_U}d shows the final $Q$-image.
	The structure seen in Fig.~\ref{fig:data_reduction_steps_Q_U}c is clearly removed.
The red solid curve in Fig.~\ref{fig:contrast_curve_hr8799} shows that the final contrast achieved is $\sim10^{-7}$ at the position of the three inner planets and a bit better at the position of planet b. 
	As illustrated by the black dashed curves in Fig.~\ref{fig:contrast_curve_hr8799}, we should be able to detect the polarization signals from the planets if they are 1\% polarized.
	The contrast reached is very close to the photon limit (purple dashed curve), that is computed as the square root of the total intensity (solid purple curve).
	Applying ADI + PCA does not increase the contrast in case of planet b, since it is at a large separation from the star and not speckle noise limited.
	At the position of the three inner planets, applying ADI + PCA has improved the contrast only by a factor of $\sim2$,	because IRDIS' dual-beam polarimetric mode already substantially suppresses speckles and we have reached the photon noise limit (i.e.~the fundamental lower contrast limit for a given dataset).
	If we would observe longer, the contrast will likely benefit more from ADI + PCA.

Reaching the contrast required to measure a $\sim1\%$ polarization signal of PZ Tel B in the data sets of 10-10-2016 (H-band) and 12-10-2016 (J-band) is less challenging.
	In the contrast curve of the H-band measurements of PZ Tel shown in Fig.~\ref{fig:contrast_curve_pztel}, the top of the companion's PSF profile (lower black solid curve) is at a contrast of $\sim6\cdot10^{-3}$.
	Hence to measure a 1\% polarized signal (black dashed curve), the contrast required is $\sim300$ times lower than for the inner three planets around HR 8799.

To create the $Q$- and $U$-images of the PZ Tel data, we omit the ADI + PCA step and instead derotate the frames after correcting the IP with the model. 
	The solid orange curve in Fig.~\ref{fig:contrast_curve_pztel} shows that we achieve a contrast of $\sim3\cdot10^{-7}$ at the position of the companion, also close to the photon limit.
	This contrast is sufficient to detect polarization signals $<0.1\%$.
	Note that because the average of the IP in Stokes $Q$ happened to be zero, the contrast curves with IP (solid green curve) and without IP (solid orange curve) are overlapping in Fig.~\ref{fig:contrast_curve_pztel}.						

\section{Results}
\label{sec:results}																				

\begin{figure}[!hbtp] 
\centering 
\includegraphics[width=\hsize]{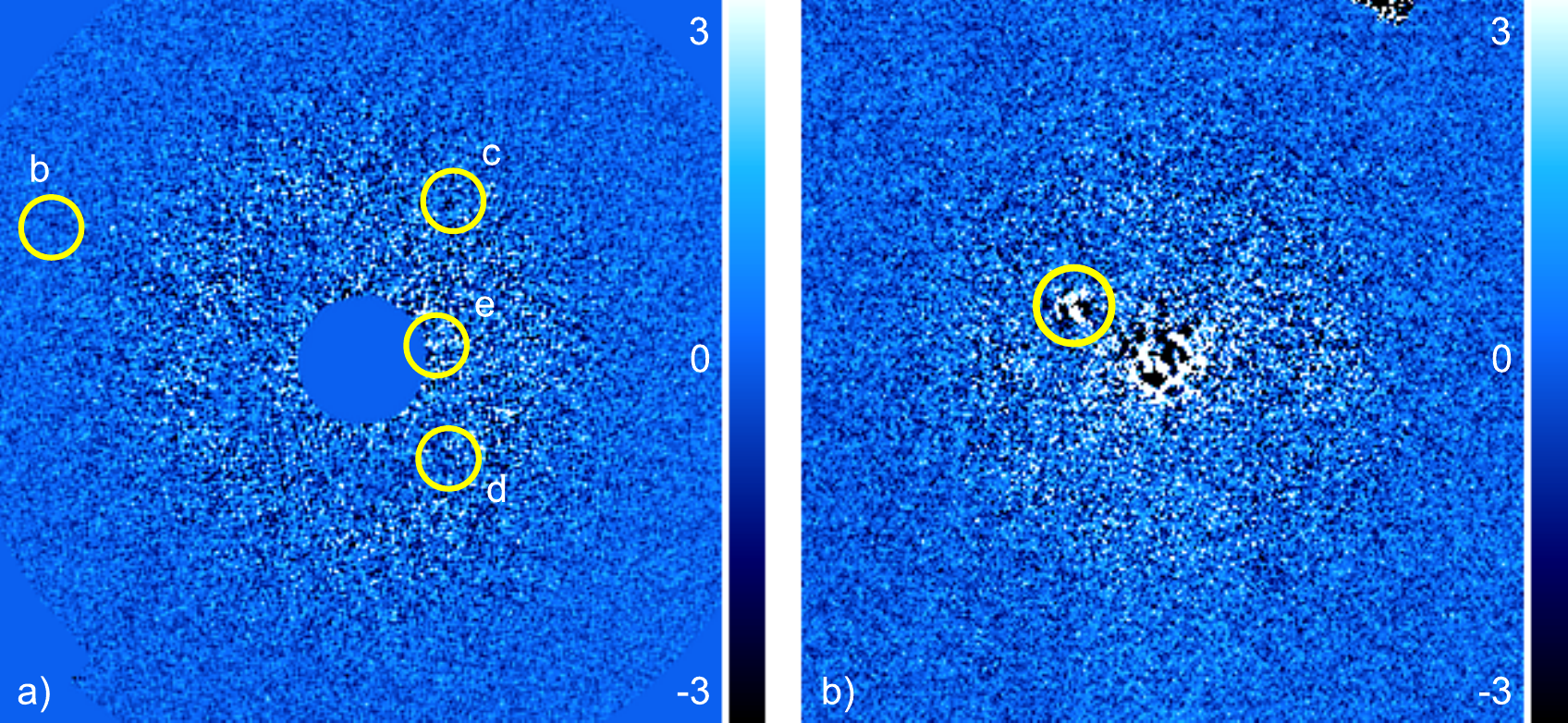} 
\caption{Final Stokes $Q$-images of the observations of HR 8799 (a) and PZ Tel (b) in H-band. By visual inspection, no polarization signals are detected.} 
\label{fig:upper_limits_hr8799_pztel} 
\end{figure}

The final Stokes $Q$-images of the H-band observations of HR 8799 and PZ Tel are shown in Fig.~\ref{fig:upper_limits_hr8799_pztel}a and \ref{fig:upper_limits_hr8799_pztel}b, respectively. 
	By visual inspection, we do not detect a polarization signal for any of the companions in our measurements. 
	Based on the contrasts achieved in Stokes $Q$ (see Figs.~\ref{fig:contrast_curve_hr8799} and \ref{fig:contrast_curve_pztel}) and $U$, we estimate preliminary $1\sigma$ upper limits on the degree of linear polarization of $\sim1\%$ and $\sim0.1\%$ for the planets of HR 8799 (H-band) and PZ Tel B (H- and J-band), respectively.
	The upper limits on the polarization of PZ Tel B would be lower based on the contrast curves alone, but is limited by the accuracy of the Mueller matrix model.
	As we will make a few improvements to the data-reduction (see Sec.~\ref{sec:discussion}), we leave the accurate determination of the polarization signals or upper limits and the interpretation of these results for future work.

\section{DISCUSSION}
\label{sec:discussion}

Although we have not detected a polarization signal, our technique of combining ADI and accurate polarimetry shows that it is possible to measure realistic polarization degrees for planetary and brown dwarf companions. 
	While for PZ Tel B we have reached a sufficiently high contrast to detect a polarization signal of a few tenths of a percent, we would need to observe the planets of HR 8799 for several nights at good seeing conditions to detect such a signal. 
	To detect polarization signals of (sub-stellar) companions without requiring an excessive amount of observing time (maximum $\sim2.5$ hr), several criteria need to be kept in mind. We would in general need (approximate numbers for the criteria are given in parentheses):
\begin{itemize}
\setlength\itemsep{0cm}
\item a brighter companion in absolute terms (H-band apparent magnitude of $\sim16$ or brighter)
\item a lower companion-to-star contrast (H-band magnitude difference of at most $\sim9$)
\item a larger separation from the central star (at least 0.6'')
\item good seeing conditions (seeing $\sim0.6''$ and coherence time $\sim4$ ms)
\item a brighter central star for good adaptive optics performance (R-band magnitude of $\sim11$ or brighter)
\end{itemize}
The first two criteria are connected: if a companion is brighter in absolute terms, a higher companion-to-star contrast is acceptable and vice versa.
	When taking all these criteria into account, only a few dozen known targets are feasible.
	To increase the chance of a detection, we should observe targets that are more likely to be polarized (see Ref.~\citenum{stolker_exopol}).
	Examples are companions with evidence for dust, hazes or patchy clouds in their atmospheres, companions that are known to rotate rapidly or that have the same spectral type and temperature as field brown dwarfs that are known to rotate rapidly (see Ref.~\citenum{milespaez_browndwarf}), companions with low surface gravities in orbits seen edge-on (higher polarization signal if rotationally flattened~\cite{dekok_exopol, stolker_exopol}) or companions that show evidence for accretion.

Since IRDIS' polarimetric mode alone (without ADI) is already so powerful for the detection of companion polarization signals, one might think to prefer the simpler field-tracking mode over the pupil-tracking mode.
	However, for close-in and/or relatively faint planets such as the planets of HR 8799, accurate determination of their total intensity (which is required to compute the degree of linear polarization) is not possible in field-tracking mode, because the PSF of the star washes out that of the planets.
	Pupil-tracking observations on the other hand allow the application of advanced ADI techniques (e.g. PCA) to effectively remove the halo of starlight and accurately determine the total intensity.
	In addition, pupil-tracking mode enables the use of the spider mask, so that the speckle noise due to the support structure of the telescope's secondary mirror is suppressed. 
	Also, because the speckles are quasi-static in pupil-tracking mode, these speckles and the IP downstream of the HWP are more effectively removed when computing the double difference.
	For long observations, applying ADI + PCA on the $Q$- and $U$-frames can possibly yield a large increase in sensitivity.
	Finally, since the planet moves over the detector during an observing sequence, flat-fielding errors are averaged out and the effect of an inconveniently located bad pixel will be limited (dithering is not yet implemented for IRDIS polarimetry). 

When performing polarimetric measurements in pupil-tracking, one should consider the effect of the parallactic rotation.	
	More parallactic rotation will be beneficial to the ADI performance and the suppression of flat-fielding errors. 
	However, when the parallactic rotation is fast, more spurious polarization signal will be created and the accuracy of the measured angle of linear polarization will diminish, because the polarization direction rotates during a single exposure with the current rotation law of the HWP (see Sec.~\ref{sec:measurement_technique}). 
	To limit these effects, one could avoid observing at the meridian or keep the exposure time and the duration of the HWP cycles short.

Before accurately computing (upper limits on) the polarization signals of the companions, we will make a few improvements to the data-reduction.
	Firstly, we will improve the centering of the frames.
	The position of the central star obtained from the star center frames differs by approximately half a pixel between the start and the end of the HR 8799 observations.
	To account for this drift, we can interpolate between the start and end coordinates of the central star and center the science frames with the interpolated coordinates.
	After that, we can further improve the centering by cross-correlating the frames.
	For the ADI + PCA data-reduction step, different PCA algorithms can be tried out.
	Most importantly, we will account for the self-subtraction of ADI + PCA, for example by using the method of fake negative planets~\cite{marois_photometry, bonnefoy_betapictorisb, zurlo_irdisphotometry}.
	Alternatively, we could use the PCA code by Ref.~\citenum{meshkat_opca} to construct for each frame to be reduced a separate stellar PSF model from only those frames where the companion PSF does not overlap with the companion PSF in the to-be-reduced frame (see~Ref.~\citenum{juanovelar_thesis}).
	Finally, we will attempt to reduce the noise in our final images by applying a matched filter, similarly to Ref.~\citenum{juanovelar_thesis}.
	When these improvements have been implemented, we will determine the degree and angle of linear polarization of the companions by performing aperture photometry on the final $I_Q$-, $I_U$-, $Q$- and $U$-frames. 
	In case of a non-detection, we will determine upper limits on the polarization by estimating the random noise in the images.
 
\section{CONCLUSION}
\label{sec:conclusions}

We have introduced a novel high-contrast imaging scheme that combines angular differential imaging (ADI) and accurate near-infrared polarimetry to characterize self-luminous giant exoplanets.
The combination of ADI and polarimetry is challenging, because the field rotation required for ADI negatively affects the polarimetric performance.
	By combining ADI and polarimetry we can characterize planets that can be directly imaged with a very high signal-to-noise ratio.
	We use the IRDIS pupil-tracking mode and combine ADI and principal component analysis to reduce speckle noise. 
	We take advantage of IRDIS' dual-beam polarimetric mode to eliminate differential effects that severely limit the polarimetric sensitivity (flat-fielding errors, differential aberrations and seeing), and thus further suppress speckle noise. 
	To correct for instrumental polarization effects, we apply a detailed Mueller matrix model that describes the telescope and instrument and that has an absolute polarimetric accuracy $\leq0.1\%$.
	As the technique is still in development, further improvements will be made in future work.

With our observing technique, we have observed the planets of HR 8799 and the (sub-stellar) companion PZ Tel B. 
	Even though by visual inspection we do not detect a polarization signal, we reach a contrast of $\sim10^{-7}$, close to the photon noise limit. 
	Based on the contrast achieved, we estimate a preliminary $1\sigma$ upper limit on the degree of linear polarization of PZ Tel B equal to $\sim0.1\%$ in H- and J-band.
	The planets of HR 8799 are much fainter however, and we estimate an upper limit of $\sim1\%$ on their degrees of linear polarization in H-band.	
	We leave the accurate determination of the polarization signals or upper limits and the interpretation of these results for future work.
	The achieved sub-percent sensitivity and accuracy show that our technique has great promise to characterize exoplanets through direct-imaging polarimetry. \\ 

\noindent\emph{The research of Frans Snik leading to these results has received funding from the European Research Council under ERC Starting Grant agreement 678194 (FALCONER).}

\bibliography{report} 

\begin{thebibliography}{10}

\bibitem{lafreniere_exoplanet}
{Lafreni{\`e}re}, D., {Jayawardhana}, R., and {van Kerkwijk}, M.~H., ``{Direct
  Imaging and Spectroscopy of a Planetary-Mass Candidate Companion to a Young
  Solar Analog},'' {\em Astrophysical Journal Letters}~{\bf 689},  L153--L156
  (2008).

\bibitem{ingraham_hr8799cd}
{Ingraham}, P., {Marley}, M.~S., {Saumon}, D., {Marois}, C., {Macintosh}, B.,
  {Barman}, T., {Bauman}, B., {Burrows}, A., {Chilcote}, J.~K., {De Rosa},
  R.~J., {Dillon}, D., {Doyon}, R., {Dunn}, J., {Erikson}, D., {Fitzgerald},
  M.~P., {Gavel}, D., {Goodsell}, S.~J., {Graham}, J.~R., {Hartung}, M.,
  {Hibon}, P., {Kalas}, P.~G., {Konopacky}, Q., {Larkin}, J.~A., {Maire}, J.,
  {Marchis}, F., {McBride}, J., {Millar-Blanchaer}, M., {Morzinski}, K.~M.,
  {Norton}, A., {Oppenheimer}, R., {Palmer}, D.~W., {Patience}, J., {Perrin},
  M.~D., {Poyneer}, L.~A., {Pueyo}, L., {Rantakyr{\"o}}, F., {Sadakuni}, N.,
  {Saddlemyer}, L., {Savransky}, D., {Soummer}, R., {Sivaramakrishnan}, A.,
  {Song}, I., {Thomas}, S., {Wallace}, J.~K., {Wiktorowicz}, S.~J., and
  {Wolff}, S.~G., ``{Gemini Planet Imager Spectroscopy of the HR 8799 Planets c
  and d},'' {\em Astrophysical Journal Letters}~{\bf 794},  L15 (2014).

\bibitem{chilcote_betapic}
{Chilcote}, J., {Barman}, T., {Fitzgerald}, M.~P., {Graham}, J.~R., {Larkin},
  J.~E., {Macintosh}, B., {Bauman}, B., {Burrows}, A.~S., {Cardwell}, A., {De
  Rosa}, R.~J., {Dillon}, D., {Doyon}, R., {Dunn}, J., {Erikson}, D., {Gavel},
  D., {Goodsell}, S.~J., {Hartung}, M., {Hibon}, P., {Ingraham}, P., {Kalas},
  P., {Konopacky}, Q., {Maire}, J., {Marchis}, F., {Marley}, M.~S., {Marois},
  C., {Millar-Blanchaer}, M., {Morzinski}, K., {Norton}, A., {Oppenheimer}, R.,
  {Palmer}, D., {Patience}, J., {Perrin}, M., {Poyneer}, L., {Pueyo}, L.,
  {Rantakyr{\"o}}, F.~T., {Sadakuni}, N., {Saddlemyer}, L., {Savransky}, D.,
  {Serio}, A., {Sivaramakrishnan}, A., {Song}, I., {Soummer}, R., {Thomas}, S.,
  {Wallace}, J.~K., {Wiktorowicz}, S., and {Wolff}, S., ``{The First H-band
  Spectrum of the Giant Planet {$\beta$} Pictoris b},'' {\em Astrophysical
  Journal Letters}~{\bf 798},  L3 (2015).

\bibitem{macintosh_erib}
{Macintosh}, B., {Graham}, J.~R., {Barman}, T., {De Rosa}, R.~J., {Konopacky},
  Q., {Marley}, M.~S., {Marois}, C., {Nielsen}, E.~L., {Pueyo}, L., {Rajan},
  A., {Rameau}, J., {Saumon}, D., {Wang}, J.~J., {Patience}, J., {Ammons}, M.,
  {Arriaga}, P., {Artigau}, E., {Beckwith}, S., {Brewster}, J., {Bruzzone}, S.,
  {Bulger}, J., {Burningham}, B., {Burrows}, A.~S., {Chen}, C., {Chiang}, E.,
  {Chilcote}, J.~K., {Dawson}, R.~I., {Dong}, R., {Doyon}, R., {Draper}, Z.~H.,
  {Duch{\^e}ne}, G., {Esposito}, T.~M., {Fabrycky}, D., {Fitzgerald}, M.~P.,
  {Follette}, K.~B., {Fortney}, J.~J., {Gerard}, B., {Goodsell}, S.,
  {Greenbaum}, A.~Z., {Hibon}, P., {Hinkley}, S., {Cotten}, T.~H., {Hung},
  L.-W., {Ingraham}, P., {Johnson-Groh}, M., {Kalas}, P., {Lafreniere}, D.,
  {Larkin}, J.~E., {Lee}, J., {Line}, M., {Long}, D., {Maire}, J., {Marchis},
  F., {Matthews}, B.~C., {Max}, C.~E., {Metchev}, S., {Millar-Blanchaer},
  M.~A., {Mittal}, T., {Morley}, C.~V., {Morzinski}, K.~M., {Murray-Clay}, R.,
  {Oppenheimer}, R., {Palmer}, D.~W., {Patel}, R., {Perrin}, M.~D., {Poyneer},
  L.~A., {Rafikov}, R.~R., {Rantakyr{\"o}}, F.~T., {Rice}, E.~L., {Rojo}, P.,
  {Rudy}, A.~R., {Ruffio}, J.-B., {Ruiz}, M.~T., {Sadakuni}, N., {Saddlemyer},
  L., {Salama}, M., {Savransky}, D., {Schneider}, A.~C., {Sivaramakrishnan},
  A., {Song}, I., {Soummer}, R., {Thomas}, S., {Vasisht}, G., {Wallace}, J.~K.,
  {Ward-Duong}, K., {Wiktorowicz}, S.~J., {Wolff}, S.~G., and {Zuckerman}, B.,
  ``{Discovery and spectroscopy of the young jovian planet 51 Eri b with the
  Gemini Planet Imager},'' {\em Science}~{\bf 350},  64--67 (2015).

\bibitem{zurlo_hr8799}
{Zurlo}, A., {Vigan}, A., {Galicher}, R., {Maire}, A.-L., {Mesa}, D.,
  {Gratton}, R., {Chauvin}, G., {Kasper}, M., {Moutou}, C., {Bonnefoy}, M.,
  {Desidera}, S., {Abe}, L., {Apai}, D., {Baruffolo}, A., {Baudoz}, P.,
  {Baudrand}, J., {Beuzit}, J.-L., {Blancard}, P., {Boccaletti}, A.,
  {Cantalloube}, F., {Carle}, M., {Cascone}, E., {Charton}, J., {Claudi},
  R.~U., {Costille}, A., {de Caprio}, V., {Dohlen}, K., {Dominik}, C.,
  {Fantinel}, D., {Feautrier}, P., {Feldt}, M., {Fusco}, T., {Gigan}, P.,
  {Girard}, J.~H., {Gisler}, D., {Gluck}, L., {Gry}, C., {Henning}, T.,
  {Hugot}, E., {Janson}, M., {Jaquet}, M., {Lagrange}, A.-M., {Langlois}, M.,
  {Llored}, M., {Madec}, F., {Magnard}, Y., {Martinez}, P., {Maurel}, D.,
  {Mawet}, D., {Meyer}, M.~R., {Milli}, J., {Moeller-Nilsson}, O., {Mouillet},
  D., {Orign{\'e}}, A., {Pavlov}, A., {Petit}, C., {Puget}, P., {Quanz}, S.~P.,
  {Rabou}, P., {Ramos}, J., {Rousset}, G., {Roux}, A., {Salasnich}, B.,
  {Salter}, G., {Sauvage}, J.-F., {Schmid}, H.~M., {Soenke}, C., {Stadler}, E.,
  {Suarez}, M., {Turatto}, M., {Udry}, S., {Vakili}, F., {Wahhaj}, Z., {Wildi},
  F., and {Antichi}, J., ``{First light of the VLT planet finder SPHERE. III.
  New spectrophotometry and astrometry of the HR 8799 exoplanetary system},''
  {\em Astronomy and Astrophysics}~{\bf 587},  A57 (2016).

\bibitem{seager_cegp}
{Seager}, S., {Whitney}, B.~A., and {Sasselov}, D.~D., ``{Photometric Light
  Curves and Polarization of Close-in Extrasolar Giant Planets},'' {\em
  Astrophysical Journal}~{\bf 540},  504 (2000).

\bibitem{stam_exo}
{Stam}, D.~M., {Hovenier}, J.~W., and {Waters}, L.~B.~F.~M., ``{Using
  polarimetry to detect and characterize Jupiter-like extrasolar planets},''
  {\em Astronomy and Astrophysics}~{\bf 428},  663 (2004).

\bibitem{sengupta_browndwarf}
{Sengupta}, S. and {Marley}, M.~S., ``{Observed Polarization of Brown Dwarfs
  Suggests Low Surface Gravity},'' {\em Astrophysical Journal Letters}~{\bf
  722},  L142--L146 (2010).

\bibitem{dekok_exopol}
{de Kok}, R.~J., {Stam}, D.~M., and {Karalidi}, T., ``{Characterizing
  Exoplanetary Atmospheres through Infrared Polarimetry},'' {\em Astrophysical
  Journal}~{\bf 741},  59 (2011).

\bibitem{stolker_exopol}
{Stolker}, T., {Min}, M., {Stam}, D.~M., {Molli{\`e}re}, P., {Dominik}, C., and
  {Waters}, R., ``{Polarized scattered light from self-luminous exoplanets},''
  {\em Astronomy and Astrophysics}  (2017).

\bibitem{sengupta_exomoons}
{Sengupta}, S. and {Marley}, M.~S., ``{Detecting Exomoons around Self-luminous
  Giant Exoplanets through Polarization},'' {\em Astrophysical Journal}~{\bf
  824},  76 (2016).

\bibitem{marley_exopol}
{Marley}, M.~S. and {Sengupta}, S., ``{Probing the physical properties of
  directly imaged gas giant exoplanets through polarization},'' {\em Monthly
  Notices of the RAS}~{\bf 417},  2874--2881 (2011).

\bibitem{bonnefoy_hr8799}
{Bonnefoy}, M., {Zurlo}, A., {Baudino}, J.~L., {Lucas}, P., {Mesa}, D.,
  {Maire}, A.-L., {Vigan}, A., {Galicher}, R., {Homeier}, D., {Marocco}, F.,
  {Gratton}, R., {Chauvin}, G., {Allard}, F., {Desidera}, S., {Kasper}, M.,
  {Moutou}, C., {Lagrange}, A.-M., {Antichi}, J., {Baruffolo}, A., {Baudrand},
  J., {Beuzit}, J.-L., {Boccaletti}, A., {Cantalloube}, F., {Carbillet}, M.,
  {Charton}, J., {Claudi}, R.~U., {Costille}, A., {Dohlen}, K., {Dominik}, C.,
  {Fantinel}, D., {Feautrier}, P., {Feldt}, M., {Fusco}, T., {Gigan}, P.,
  {Girard}, J.~H., {Gluck}, L., {Gry}, C., {Henning}, T., {Janson}, M.,
  {Langlois}, M., {Madec}, F., {Magnard}, Y., {Maurel}, D., {Mawet}, D.,
  {Meyer}, M.~R., {Milli}, J., {Moeller-Nilsson}, O., {Mouillet}, D., {Pavlov},
  A., {Perret}, D., {Pujet}, P., {Quanz}, S.~P., {Rochat}, S., {Rousset}, G.,
  {Roux}, A., {Salasnich}, B., {Salter}, G., {Sauvage}, J.-F., {Schmid}, H.~M.,
  {Sevin}, A., {Soenke}, C., {Stadler}, E., {Turatto}, M., {Udry}, S.,
  {Vakili}, F., {Wahhaj}, Z., and {Wildi}, F., ``{First light of the VLT planet
  finder SPHERE. IV. Physical and chemical properties of the planets around
  HR8799},'' {\em Astronomy and Astrophysics}~{\bf 587},  A58 (2016).

\bibitem{milespaez_browndwarf}
{Miles-P{\'a}ez}, P.~A., {Zapatero Osorio}, M.~R., {Pall{\'e}}, E., and
  {Pe{\~n}a Ram{\'{\i}}rez}, K., ``{Linear polarization of rapidly rotating
  ultracool dwarfs},'' {\em Astronomy and Astrophysics}~{\bf 556},  A125
  (2013).

\bibitem{juanovelar_thesis}
{Juan Ovelar}, M.~d., {\em {Imaging polarimetry for the characterisation of
  exoplanets and protoplanetary discs: scientific and technical challenges}},
  {Ph.D.} {T}hesis, Leiden Observatory, Faculty of Science, Leiden University
  (2013).

\bibitem{jensen_padi}
{Jensen-Clem}, R., {Millar-Blanchaer}, M., {Mawet}, D., {Graham}, J.~R.,
  {Wallace}, J.~K., {Macintosh}, B., {Hinkley}, S., {Wiktorowicz}, S.~J.,
  {Perrin}, M.~D., {Marley}, M.~S., {Fitzgerald}, M.~P., {Oppenheimer}, R.,
  {Ammons}, S.~M., {Rantakyr{\"o}}, F.~T., and {Marchis}, F., ``{Point Source
  Polarimetry with the Gemini Planet Imager: Sensitivity Characterization with
  T5.5 Dwarf Companion HD 19467 B},'' {\em Astrophysical Journal}~{\bf 820},
  111 (2016).

\bibitem{beuzit_sphere}
{Beuzit}, J.-L., {Feldt}, M., {Dohlen}, K., {Mouillet}, D., {Puget}, P.,
  {Wildi}, F., {Abe}, L., {Antichi}, J., {Baruffolo}, A., {Baudoz}, P.,
  {Boccaletti}, A., {Carbillet}, M., {Charton}, J., {Claudi}, R., {Downing},
  M., {Fabron}, C., {Feautrier}, P., {Fedrigo}, E., {Fusco}, T., {Gach}, J.-L.,
  {Gratton}, R., {Henning}, T., {Hubin}, N., {Joos}, F., {Kasper}, M.,
  {Langlois}, M., {Lenzen}, R., {Moutou}, C., {Pavlov}, A., {Petit}, C.,
  {Pragt}, J., {Rabou}, P., {Rigal}, F., {Roelfsema}, R., {Rousset}, G.,
  {Saisse}, M., {Schmid}, H.-M., {Stadler}, E., {Thalmann}, C., {Turatto}, M.,
  {Udry}, S., {Vakili}, F., and {Waters}, R., ``{SPHERE: a 'Planet Finder'
  instrument for the VLT},'' {\em Proceedings of SPIE}~{\bf 7014} (2008).

\bibitem{dohlen_irdis}
{Dohlen}, K., {Langlois}, M., {Saisse}, M., {Hill}, L., {Origne}, A.,
  {Jacquet}, M., {Fabron}, C., {Blanc}, J.-C., {Llored}, M., {Carle}, M.,
  {Moutou}, C., {Vigan}, A., {Boccaletti}, A., {Carbillet}, M., {Mouillet}, D.,
  and {Beuzit}, J.-L., ``{The infra-red dual imaging and spectrograph for
  SPHERE: design and performance},'' {\em Proceedings of SPIE}~{\bf 7014},  3
  (2008).

\bibitem{langlois_irdis}
{Langlois}, M., {Dohlen}, K., {Vigan}, A., {Zurlo}, A., {Moutou}, C., {Schmid},
  H.~M., {Mili}, J., {Beuzit}, J.-L., {Boccaletti}, A., {Carle}, M.,
  {Costille}, A., {Dorn}, R., {Gluck}, L., {Hubin}, N., {Feldt}, M., {Kasper},
  M., {Lizon}, L., {Madec}, F., {Le Mignant}, D., {Mouillet}, D., {Puget},
  J.-P., {Sauvage}, J.-F., and {Wildi}, F., ``{High contrast polarimetry in the
  infrared with SPHERE on the VLT},'' {\em Proceedings of SPIE}~{\bf 9147},  1
  (2014).

\bibitem{deboer_irdis}
{de Boer}, J. {\em \emph{et al.} Astronomy and Astrophysics \emph{(in prep.)}}
  .

\bibitem{marois_hr8799}
{Marois}, C., {Macintosh}, B., {Barman}, T., {Zuckerman}, B., {Song}, I.,
  {Patience}, J., {Lafreni{\`e}re}, D., and {Doyon}, R., ``{Direct Imaging of
  Multiple Planets Orbiting the Star HR 8799},'' {\em Science}~{\bf 322},
  1348-- (2008).

\bibitem{marois_hr8799e}
{Marois}, C., {Zuckerman}, B., {Konopacky}, Q.~M., {Macintosh}, B., and
  {Barman}, T., ``{Images of a fourth planet orbiting HR 8799},'' {\em
  Nature}~{\bf 468},  1080--1083 (2010).

\bibitem{biller_pztel}
{Biller}, B.~A., {Liu}, M.~C., {Wahhaj}, Z., {Nielsen}, E.~L., {Close}, L.~M.,
  {Dupuy}, T.~J., {Hayward}, T.~L., {Burrows}, A., {Chun}, M., {Ftaclas}, C.,
  {Clarke}, F., {Hartung}, M., {Males}, J., {Reid}, I.~N., {Shkolnik}, E.~L.,
  {Skemer}, A., {Tecza}, M., {Thatte}, N., {Alencar}, S.~H.~P., {Artymowicz},
  P., {Boss}, A., {de Gouveia Dal Pino}, E., {Gregorio-Hetem}, J., {Ida}, S.,
  {Kuchner}, M.~J., {Lin}, D., and {Toomey}, D., ``{The Gemini NICI
  Planet-finding Campaign: Discovery of a Close Substellar Companion to the
  Young Debris Disk Star PZ Tel},'' {\em Astrophysical Journal Letters}~{\bf
  720},  L82--L87 (Sept. 2010).

\bibitem{mugrauer_pztel}
{Mugrauer}, M., {Vogt}, N., {Neuh{\"a}user}, R., and {Schmidt}, T.~O.~B.,
  ``{Direct detection of a substellar companion to the young nearby star PZ
  Telescopii},'' {\em Astronomy and Astrophysics}~{\bf 523},  L1 (Nov. 2010).

\bibitem{oppenheimer_hr8799}
{Oppenheimer}, B.~R., {Baranec}, C., {Beichman}, C., {Brenner}, D., {Burruss},
  R., {Cady}, E., {Crepp}, J.~R., {Dekany}, R., {Fergus}, R., {Hale}, D.,
  {Hillenbrand}, L., {Hinkley}, S., {Hogg}, D.~W., {King}, D., {Ligon}, E.~R.,
  {Lockhart}, T., {Nilsson}, R., {Parry}, I.~R., {Pueyo}, L., {Rice}, E.,
  {Roberts}, J.~E., {Roberts}, Jr., L.~C., {Shao}, M., {Sivaramakrishnan}, A.,
  {Soummer}, R., {Truong}, T., {Vasisht}, G., {Veicht}, A., {Vescelus}, F.,
  {Wallace}, J.~K., {Zhai}, C., and {Zimmerman}, N., ``{Reconnaissance of the
  HR 8799 Exosolar System. I. Near-infrared Spectroscopy},'' {\em Astrophysical
  Journal}~{\bf 768},  24 (2013).

\bibitem{ginski_companionorbits}
{Ginski}, C., {Schmidt}, T.~O.~B., {Mugrauer}, M., {Neuh{\"a}user}, R., {Vogt},
  N., {Errmann}, R., and {Berndt}, A., ``{Astrometric follow-up observations of
  directly imaged sub-stellar companions to young stars and brown dwarfs},''
  {\em Monthly Notices of the RAS}~{\bf 444},  2280--2302 (Nov. 2014).

\bibitem{maire_pztel}
{Maire}, A.-L., {Bonnefoy}, M., {Ginski}, C., {Vigan}, A., {Messina}, S.,
  {Mesa}, D., {Galicher}, R., {Gratton}, R., {Desidera}, S., {Kopytova}, T.~G.,
  {Millward}, M., {Thalmann}, C., {Claudi}, R.~U., {Ehrenreich}, D., {Zurlo},
  A., {Chauvin}, G., {Antichi}, J., {Baruffolo}, A., {Bazzon}, A., {Beuzit},
  J.-L., {Blanchard}, P., {Boccaletti}, A., {de Boer}, J., {Carle}, M.,
  {Cascone}, E., {Costille}, A., {De Caprio}, V., {Delboulb{\'e}}, A.,
  {Dohlen}, K., {Dominik}, C., {Feldt}, M., {Fusco}, T., {Girard}, J.~H.,
  {Giro}, E., {Gisler}, D., {Gluck}, L., {Gry}, C., {Henning}, T., {Hubin}, N.,
  {Hugot}, E., {Jaquet}, M., {Kasper}, M., {Lagrange}, A.-M., {Langlois}, M.,
  {Le Mignant}, D., {Llored}, M., {Madec}, F., {Martinez}, P., {Mawet}, D.,
  {Milli}, J., {M{\"o}ller-Nilsson}, O., {Mouillet}, D., {Moulin}, T.,
  {Moutou}, C., {Orign{\'e}}, A., {Pavlov}, A., {Petit}, C., {Pragt}, J.,
  {Puget}, P., {Ramos}, J., {Rochat}, S., {Roelfsema}, R., {Salasnich}, B.,
  {Sauvage}, J.-F., {Schmid}, H.~M., {Turatto}, M., {Udry}, S., {Vakili}, F.,
  {Wahhaj}, Z., {Weber}, L., and {Wildi}, F., ``{First light of the VLT planet
  finder SPHERE. II. The physical properties and the architecture of the young
  systems PZ Telescopii and HD 1160 revisited},'' {\em Astronomy and
  Astrophysics}~{\bf 587},  A56 (Mar. 2016).

\bibitem{marois_adi}
{Marois}, C., {Lafreni{\`e}re}, D., {Doyon}, R., {Macintosh}, B., and {Nadeau},
  D., ``{Angular Differential Imaging: A Powerful High-Contrast Imaging
  Technique},'' {\em Astrophysical Journal}~{\bf 641},  556--564 (2006).

\bibitem{vanholstein_irdismodel}
{van Holstein}, R.~G. {\em \emph{et al.} Astronomy and Astrophysics \emph{(in
  prep.)}} .

\bibitem{bagnulo_spectropolarimetry}
{Bagnulo}, S., {Landolfi}, M., {Landstreet}, J.~D., {Landi Degl'Innocenti}, E.,
  {Fossati}, L., and {Sterzik}, M., ``{Stellar Spectropolarimetry with Retarder
  Waveplate and Beam Splitter Devices},'' {\em Publications of the ASP}~{\bf
  121},  993--1015 (2009).

\bibitem{canovas_data}
{Canovas}, H., {Rodenhuis}, M., {Jeffers}, S.~V., {Min}, M., and {Keller},
  C.~U., ``{Data-reduction techniques for high-contrast imaging polarimetry.
  Applications to ExPo},'' {\em Astronomy and Astrophysics}~{\bf 531},  A102
  (2011).

\bibitem{amara_pynpoint}
{Amara}, A. and {Quanz}, S.~P., ``{PYNPOINT: an image processing package for
  finding exoplanets},'' {\em Monthly Notices of the RAS}~{\bf 427},  948--955
  (2012).

\bibitem{soummer_klip}
{Soummer}, R., {Pueyo}, L., and {Larkin}, J., ``{Detection and Characterization
  of Exoplanets and Disks Using Projections on Karhunen-Lo{\`e}ve
  Eigenimages},'' {\em Astrophysical Journal Letters}~{\bf 755},  L28 (2012).

\bibitem{meshkat_opca}
{Meshkat}, T., {Kenworthy}, M.~A., {Quanz}, S.~P., and {Amara}, A.,
  ``{Optimized Principal Component Analysis on Coronagraphic Images of the
  Fomalhaut System},'' {\em Astrophysical Journal}~{\bf 780},  17 (2014).

\bibitem{sphere_manual}
{Wahhaj}, Z., {Girard}, J., {Milli}, J., {Vigan}, M., {van den Ancker}, M.,
  {Boffin}, H., {Mouillet}, D., {Mawet}, D., {Kasper}, R., {Siebenmorgem}, R.,
  and {SPHERE Consortium}, ``{SPHERE User Manual Issue P100 Phase 2},'' {\em
  ESO}  (2017).

\bibitem{amara_pynpointcode}
{Amara}, A., {Quanz}, S.~P., and {Akeret}, J., ``{PynPoint code for exoplanet
  imaging},'' {\em Astronomy and Computing}~{\bf 10},  107--115 (2015).

\bibitem{marois_photometry}
{Marois}, C., {Macintosh}, B., and {V{\'e}ran}, J.-P., ``{Exoplanet imaging
  with LOCI processing: photometry and astrometry with the new SOSIE
  pipeline},'' {\em Proceedings of SPIE}~{\bf 7736},  77361J (2010).

\bibitem{bonnefoy_betapictorisb}
{Bonnefoy}, M., {Lagrange}, A.-M., {Boccaletti}, A., {Chauvin}, G., {Apai}, D.,
  {Allard}, F., {Ehrenreich}, D., {Girard}, J.~H.~V., {Mouillet}, D., {Rouan},
  D., {Gratadour}, D., and {Kasper}, M., ``{High angular resolution detection
  of {$\beta$} Pictoris b at 2.18 {$\mu$}m},'' {\em Astronomy and
  Astrophysics}~{\bf 528},  L15 (Apr. 2011).

\bibitem{zurlo_irdisphotometry}
{Zurlo}, A., {Vigan}, A., {Mesa}, D., {Gratton}, R., {Moutou}, C., {Langlois},
  M., {Claudi}, R.~U., {Pueyo}, L., {Boccaletti}, A., {Baruffolo}, A.,
  {Beuzit}, J.-L., {Costille}, A., {Desidera}, S., {Dohlen}, K., {Feldt}, M.,
  {Fusco}, T., {Henning}, T., {Kasper}, M., {Martinez}, P., {Moeller-Nilsson},
  O., {Mouillet}, D., {Pavlov}, A., {Puget}, P., {Sauvage}, J.-F., {Turatto},
  M., {Udry}, S., {Vakili}, F., {Waters}, R., and {Wildi}, R.~F.,
  ``{Performance of the VLT Planet Finder SPHERE. I. Photometry and astrometry
  precision with IRDIS and IFS in laboratory},'' {\em Astronomy and
  Astrophysics}~{\bf 572},  A85 (Dec. 2014).

\bibitem{deboer_diskcandidates}
{de Boer}, J., {Salter}, G., {Benisty}, M., {Vigan}, A., {Boccaletti}, A.,
  {Pinilla}, P., {Ginski}, C., {Juhasz}, A., {Maire}, A.-L., {Messina}, S.,
  {Desidera}, S., {Cheetham}, A., {Girard}, J.~H., {Wahhaj}, Z., {Langlois},
  M., {Bonnefoy}, M., {Beuzit}, J.-L., {Buenzli}, E., {Chauvin}, G., {Dominik},
  C., {Feldt}, M., {Gratton}, R., {Hagelberg}, J., {Isella}, A., {Janson}, M.,
  {Keller}, C.~U., {Lagrange}, A.-M., {Lannier}, J., {Menard}, F., {Mesa}, D.,
  {Mouillet}, D., {Mugrauer}, M., {Peretti}, S., {Perrot}, C., {Sissa}, E.,
  {Snik}, F., {Vogt}, N., {Zurlo}, A., and {SPHERE Consortium}, ``{Multiple
  rings in the transition disk and companion candidates around RX J1615.3-3255.
  High contrast imaging with VLT/SPHERE},'' {\em Astronomy and
  Astrophysics}~{\bf 595},  A114 (Nov. 2016).

\end{thebibliography}
\bibliographystyle{spiebib} 

\end{document}